\documentclass[journal]{IEEEtran}
\usepackage{url}
\usepackage[dvips]{graphicx}
\usepackage[cmex10]{amsmath}
\usepackage{multirow}
\usepackage{array}
\usepackage{threeparttable}
\ifCLASSINFOpdf
\else
\fi
\begin{document}
\title{Influences of pump transitions on thermal effects of multi-kilowatt thulium-doped fiber lasers}
\author{Jianlong~Yang, Yao~Wang, Yulong~Tang, and Jianqiu~Xu
\thanks{This work was supported by the National Natural Science Foundation of China (Nos. 61275136, 11121504), the Key National Natural Science Foundation of China (No. 61138006).}
\thanks{The authors are with the Key Laboratory for Laser Plasmas (Ministry of Education) and the Department of Physics and Astronomy, Shanghai Jiao Tong University, Shanghai 200240, China (email: nicholasyjl@sjtu.edu.cn)}}
\markboth{}
{Yang \MakeLowercase{\textit{et al.}}: Thermal effects in kilowatt thulium-doped fiber lasers}
\maketitle
\begin{abstract}
Thermal effects are critical constrains for developing high-power thulium-doped fiber lasers (TDFLs). In this paper, we numerically investigate the lasing and thermal characteristics of the TDFLs under different pump transitions. Our results show, the widely-used pump transition $^3H_6\rightarrow^3H_4$, taking advantages of high-power high-efficiency laser diodes at $\sim$0.8 $\mu$m, may not be a superior choice for directly outputting multi-kilowatt at 2 $\mu$m because of severe thermal problems. Meanwhile, using other pump transitions resulting 2-$\mu$m emissions, especially the in-band pump transition $^3H_6\rightarrow^3F_4$, will decrease the generated heat to a large extent. By optimizing the power filling factor of the gain fiber, we find a 2-$\mu$m TDFL cladding-pumped at 1.9 $\mu$m will lead to the laser slope efficiency close to its quantum efficiency (95\%). The induced ultra-low quantum defect would be of great importance for power scaling. We thus propose tandem-pumped TDFLs for reducing the heat at high powers and discuss the related issues of design. Besides, we also explore the differences of the thermal characteristics between laser and superfluorescent operations, which will contribute to deepening the understanding of the thermal effects in high-power thulium-doped fiber amplifiers.
\end{abstract}
\IEEEpeerreviewmaketitle

\section{Introduction}
\indent In recent years, the importance of thulium-doped fiber lasers (TDFLs) at 2 $\mu$m has been kept growing both in research area and laser industry \cite{carter09}, since their output power reached 1 kilowatt \cite{ehrenreich10}, the second place among fiber lasers. Comparing to their outstanding counterparts at 1 $\mu$m, ytterbium-doped fiber lasers (YDFLs), TDFLs operate at a widely tunable wavelength range from 1810 nm to 2200 nm \cite{li13,li14} covering the water absorption peak at 1.94 $\mu$m and part of atmospheric transmission window, which is favourable for many medicinal and military applications \cite{scott09,wang14,gebhardt14}. Besides, operating at longer wavelengths and the so-called eye-safe waveband will increase the thresholds of nonlinear effects \cite{richardson10} and the level of the permissible exposure, respectively.\\
\indent However, the development of high-power TDFLs is largely hindered by severe thermal effects because the heat load of the TDFLs is $8\times$ larger than that of the YDFLs \cite{goodno11}. Except some cases making use of the heat generated in fiber, such as suppression of stimulated Brillouin scattering \cite{jeong07,kovalev06}, thermal effects are mostly detrimental. Not only those taking effects at very high power outputs, such as thermal lens, melting and rupture of silica \cite{dawson08,brown01}, insufficient cooling of any fiber lasers would lead to decrease of efficiency and output power, increase of lasing threshold, and wavelength redshift \cite{brilliant01,huang14}. Besides, it would also result in the damages of fiber lasers by burning of polymer coating or fiber fuse effect \cite{jackson07,hand88} even at hundred-watt level and below.\\
\indent Different approaches has been employed to eliminate the heat generated in fiber lasers, including various designs of convective cooling \cite{moulton09,fan11} and gain fiber \cite{huang14,lapointe09},  different pump strategies \cite{pask95,wang04}, and multiple laser configurations \cite{ehrenreich10}. For example, the realization of the first kilowatt TDFL was based on the usage of all-glass fibers to avoid the burning of polymer coating \cite{ehrenreich10}. However, the endeavours on cooling and laser design can not truly settle this problem, because the heat is primarily originated from quantum defect and non-radiative decay \cite{dawson08}. The revolutionary technique that push the outputs of YDFLs to 10 kilowatt is tandem-pumping \cite{stiles09}, which essentially was employed to decrease quantum defect at high powers.\\
\indent To decrease the quantum defect, pump wavelengths close to lasing wavelengths should be employed. For the TDFLs, the pump transitions $^3H_6\rightarrow^3H_4$, $^3H_6\rightarrow^3H_5$, and $^3H_6\rightarrow^3F_4$, corresponding to several pump wavelengths, can be used to output 2-$\mu$m laser \cite{jackson99}. In this paper, we analyze their lasing performances and thermal characteristics at multi-kilowatt level by employing an experimental-validated model. Based on the numerical results, we propose tandem-pumped TDFLs to minimize the heat generation at high powers. We arrange the paper as follows. In Section \uppercase\expandafter{\romannumeral2}, we briefly summarize the characteristics of the different pump transitions. In Section \uppercase\expandafter{\romannumeral3}, we establish our theoretical model and validate it by the comparison with experimental results. In Section \uppercase\expandafter{\romannumeral4}, we demonstrate the numerical results and discuss their implications. We propose the design of the tandem-pumped TDFLs in Section \uppercase\expandafter{\romannumeral5} and drawn our conclusions in Section \uppercase\expandafter{\romannumeral6}.
\section{Characterization of different pump transitions}
\indent The pump transitions $^3H_6\rightarrow^3H_4$, $^3H_6\rightarrow^3H_5$, and $^3H_6\rightarrow^3F_4$ correspond to three absorption wavebands of thulium cations peaked at $\sim$0.8 $\mu$m, $\sim$1.2 $\mu$m, and $\sim$1.65 $\mu$m, respectively \cite{jackson99}. However, except the absorption peak of $^3H_6\rightarrow^3H_4$ coincidentally locates at the wavelengths of high-efficiency laser diodes (LDs) at $\sim$0.8 $\mu$m, it is difficult to find suitable sources for efficient excitation at the absorption peaks of other pump transitions. The pump sources emit at these wavelength-peaks are usually of low efficiency and difficult to scale up to high power, such as Raman lasers and color center lasers. Fortunately, these absorption wavebands all span over very broad spectral ranges, so various light sources can be employed as the even through the absorption cross sections are small. The related experiments can be found in \cite{jackson99,tsang04} and references therein, we are only interested here in the pump sources with fiber-based structures and having been fully exploited to deliver hundreds-of-watt output powers, because of their potential for realizing monolithic fiber lasers and capacity to provide abundant pump powers through well-developed fiber pump combiners for achieving multi-kilowatt 2-$\mu$m outputs.\\
\begin{table}[tbh]
\caption{Characteristics of different pump transitions}
\begin{center}
\begin{threeparttable}
\begin{tabular}{p{2cm}|m{1.6cm}m{1.6cm}m{0.8cm}m{0.8cm}}
\hline
\hline
Pump transition & $^3H_6\rightarrow^3H_4$ & $^3H_6\rightarrow^3H_5$ & \multicolumn{2}{c}{$^3H_6\rightarrow^3F_4$}\\
\hline
Wavelength ($\mu$m) & 0.8 & 1 & 1.6 & 1.9 \\
$\sigma_a$ ($10^{-25}$m$^2$) & 8.5 & 1.1 & 1.6 & 0.6\\
$\alpha$ ($10^{-3}$m$^{-1}$) &2.5&1.2&1.1&11.5\\
$\eta_{laser}$ (\%) & 53.2 & 37 & 72 & 90.2 \\
Pump source & LD & YDFL & EYFL & TDFL\\
\hline
\hline
\end{tabular}
\end{threeparttable}
\end{center}
\end{table}
\indent Table \uppercase\expandafter{\romannumeral1} provides basic characteristics of different pump transitions at the wavelengths that coincide with available high-power fiber-based pump sources. EYFL is the abbreviation of erbium/ytterbium-codoped fiber laser. The parameters of stimulated absorption cross section $\sigma_a$ and fiber attenuation $\alpha$ at these wavelengths are taken from \cite{jackson99,mitschke09,jackson98} and the corresponding laser slope efficiency $\eta_{laser}$ are taken from representative experimental works \cite{ehrenreich10,hanna90,shen06,creeden14}. The pump transition $^3H_6\rightarrow^3H_4$ at $\sim$0.8 $\mu$m has a very large absorption cross section, which enables its efficiency of absorption when using cladding-pumped low-brightness LDs. For example, commercial available thulium-doped fibers (TDFs) usually have a cladding absorption of $2-3$ dB/m, so the TDFLs can be efficiently operated at a gain fiber length of $\sim$5 m. However, strong pump absorption will induce substantial heat generated at the injection end of the gain fiber. A solution is to use low doping concentrations \cite{huang14} but this method will reduce the efficiency of the famous ``two-for-one'' cross relaxation (CR) process \cite{jackson07} and thus affect the $\eta_{laser}$ of the TDFLs. The emission wavelengths of high-power YDFLs lies at the edge of the $^3H_6\rightarrow^3H_5$ absorption waveband. However, the rather small absorption cross sections can be compensated by the high-brightness of fiber lasers as the pump sourceS. The disadvantages of this pump transition are its low slope efficiency related to the abundant transition processes, such as excited-state absorption (ESA) and CR, and a possible photodarken effect \cite{broer93}. Although the origin of the photodarken effect has not been clearly understood, it was usually be related to high excitation intensity \cite{peretti10}. So this effect may be insignificant in cladding-pumped TDFLs  at 1 $\mu$m because of the wake absorption. For the pump transition $^3H_6\rightarrow^3F_4$, EYFLs at $\sim$1.6 $\mu$m and TDFLs at $\sim$1.9 $\mu$m can be employed. High slope efficiencies are usually achieved under this pump transition, because it has a simple ``two-level'' energy diagram similar to that of YDFLs and very high quantum efficiency. Especially when using 1.9-$\mu$m TDFLs as the pump sources, a slope efficiency of 90.2\% has been acquired \cite{creeden14}. A quantum defect of less than 10\% will be important for the heat management at high-powers, which has be justified in the development of YDFLs. Same as that at $\sim$1 $\mu$m, the absorption cross sections at $\sim$1.6 $\mu$m and $\sim$1.9 $\mu$m are relatively small, but it will not be a problem for their high-brightness pump sources. Special attention should be paid to the large fiber attenuation at $\sim$1.9 $\mu$m, which implies a long gain fiber is unfavorable for the lasers operating at these wavelengths.
\section{theoretical modeling}
\subsection{Rate-equation model}
The simplified energy diagram of Tm$^{3+}$ cations and all involved transitions are demonstrated in Fig.~1, dashed lines was used to separate the processes related to different pump transitions. The black, blue, brown, green, and red arrows represent the stimulated absorption of the pump, ESA, CR and energy transfer upconversion (ETU), non-radiative decay, and the stimulated emission of the 2-$\mu$m laser, respectively. The illustration of each process will be given below.
\begin{figure*}[htb]
\begin{center}
\includegraphics[scale=0.32]{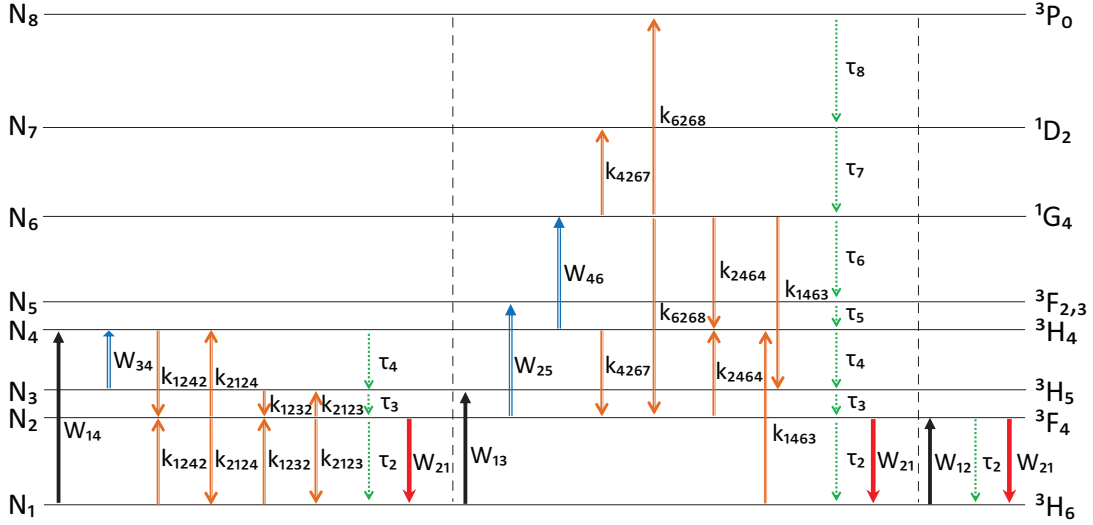}
\caption{Energy diagrams of the Tm$^{3+}$ cations, dashed lines was used to separate the processes related to different pump transitions.}
\end{center}
\end{figure*}
\subsubsection{$^3H_6\rightarrow^3H_4$}
The laser behaviors under this pump transition can be described by a set of nonlinear partial differential equations: 
\begin{align}
\frac{\partial N_4}{\partial t}&=W_{14}N_1+W_{34}N_3-k_{1242}N_1N_4+k_{2124}N_2^2-\frac{N_4}{\tau_4},\\
\frac{\partial N_3}{\partial t}&=-W_{34}N_3-k_{1232}N_1N_3+k_{2123}N_2^2+\frac{N_4}{\tau_4}-\frac{N_3}{\tau_3},\\
\frac{\partial N_2}{\partial t}&=2k_{1242}N_1N_4-2k_{2124}N_2^2+2k_{1232}N_1N_3-W_{21}N_2\nonumber\\
&+\frac{N_3}{\tau_3}-\frac{N_2}{\tau_2}-2k_{2123}N_2^2,\\
N_1&=N_{0}-N_2-N_3-N_4,
\end{align}
\begin{align}
\pm \frac{\partial P^{\pm}_p}{\partial z}&+\frac{\partial P^{\pm}_p}{v_g\partial t}=-\Gamma_p\sigma_{14}N_1P^{\pm}_p-\alpha_pP^{\pm}_p,\\
\pm \frac{\partial P^{\pm}_s}{\partial z}&+\frac{\partial P^{\pm}_s}{v_g\partial t}=\Gamma_s[\sigma_{21}N_2-\sigma_{34}N_3]P^{\pm}_s-\alpha_sP^{\pm}_s+\Gamma_s\frac{2hc^2}{\lambda_s^3}\nonumber\\
&\times \sigma_{21}N_2\Delta\lambda_{ASE},
\end{align}
where $N_0$ is the ion concentration of the Tm$^{3+}$ cations. $N_1$, $N_2$, $N_3$, $N_4$ are the population densities on the energy level $^3H_6$, $^3F_4$, $^3H_5$, and $^3H_4$, respectively. $W_{14}$ is the stimulated absorption coefficient of the pump and $W_{21}$ is the stimulated emission coefficient of the laser. $W_{34}$ is the ESA coefficient of the laser from $^3H_5$ to $^3H_4$. $\tau_2$, $\tau_3$, $\tau_4$ are the life times of the excited ions on the energy level $^3F_4$, $^3H_5$, and $^3H_4$, respectively. $k_{1242}$ is the coefficient of the CR process $^3H_6\rightarrow^3F_4\leftarrow^3H_4$ and $k_{1232}$ is the coefficient of the CR process $^3H_6\rightarrow^3F_4\leftarrow^3H_5$. $k_{2124}$ and $k_{2123}$ are the coefficients of the ETU process $^3F_4\rightarrow^3H_6\cdots^3F_4\rightarrow^3H_4$ and $^3F_4\rightarrow^3H_6\cdots^3F_4\rightarrow^3H_5$, respectively. $P_p^{\pm}$ and $P_s^{\pm}$ are the optical powers of the pump and the laser, respectively. The superscript $+$ represents the forward propagation direction and $-$ represents the backward propagation direction. $v_g$ is the group velocity. $\Gamma_p$ is the power filling factor of the pump. For cladding pump, $\Gamma_p=A_{core}/A_{clad}$, where $A_{core}$ and $A_{clad}$ are the effective areas of the fiber core and inner cladding, respectively. The $	\Gamma_p$ for core pump and the power filling factor of the laser $\Gamma_s$ approximate 1 and can be precisely deduced by following the method in \cite{yang14}. $\sigma_{14}$ and $\sigma_{34}$ are the absorption cross section of the pump and the transition $^3H_5\rightarrow^3H_4$, respectively. $\sigma_{21}$ is the emission cross section of the laser. $h$ is the Planck constant and $c$ is the velocity of light in vacuum. $\alpha_p$ and $\alpha_s$ are the fiber attenuation of the pump and the laser, respectively. $\Delta\lambda_{ASE}$ is the bandwidth of the amplified spontoneous emission (ASE) \cite{yang14}. $\lambda_s$ is the central wavelength of the laser.
\subsubsection{$^3H_6\rightarrow^3H_5$}
\begin{align}
\frac{\partial N_8}{\partial t}&=k_{6268}N_6^2-\frac{N_8}{\tau_8},\\
\frac{\partial N_7}{\partial t}&=k_{4267}N_4N_6+\frac{N_8}{\tau_8}-\frac{N_7}{\tau_7},\\
\frac{\partial N_6}{\partial t}&=W_{46}N_4-k_{4267}N_4N_6-2k_{6268}N_6^2-k_{2464}N_2N_6\nonumber\\
&-k_{1463}N_1N_6+\frac{N_7}{\tau_7}-\frac{N_6}{\tau_6},\\
\frac{\partial N_5}{\partial t}&=W_{25}N_2+\frac{N_6}{\tau_6}-\frac{N_5}{\tau_5},\\
\frac{\partial N_4}{\partial t}&=-W_{46}N_4-k_{4267}N_4N_6+2k_{2464}N_2N_6+\frac{N_5}{\tau_5}\nonumber \\
&+k_{1463}N_1N_6-\frac{N_4}{\tau_4},\\
\frac{\partial N_3}{\partial t}&=W_{13}N_1+k_{1463}N_1N_6+\frac{N_4}{\tau_4}-\frac{N_3}{\tau_3},\\
\frac{\partial N_2}{\partial t}&=-W_{25}N_2+k_{4267}N_4N_6+k_{6268}N_6^2-W_{21}N_2\nonumber \\
&+\frac{N_3}{\tau_3}-\frac{N_2}{\tau_2}-k_{2464}N_2N_6,
\end{align}
\begin{align}
N_1&=N_{0}-N_2-N_3-N_4-N_5-N_6-N_7-N_8,\\
\pm \frac{\partial P^{\pm}_p}{\partial z}&+\frac{\partial P^{\pm}_p}{v_g\partial t}=\Gamma_p[-\sigma_{13}N_1-\sigma_{25}N_2-\sigma_{46}N_4]P^{\pm}_p-\nonumber\\
&\alpha_pP^{\pm}_p,\\
\pm \frac{\partial P^{\pm}_s}{\partial z}&+\frac{\partial P^{\pm}_s}{v_g\partial t}=\Gamma_s\sigma_{21}N_2P^{\pm}_s-\alpha_sP^{\pm}_s+\Gamma_s\frac{2hc^2}{\lambda_s^3}\sigma_{21}\nonumber\\
&\times N_2\Delta\lambda_{ASE},
\end{align}
where $N_5$, $N_6$, $N_7$, and $N_8$ are the population density of $^3F_{2,3}$, $^1G_4$, $^1D_2$, and $^3P_0$, respectively, and $\tau_i\ (i=5,6,7,8)$ are their lifetimes. $W_{13}$ is the stimulated absorption coefficient of the pump. The pump ESA $^3F_4\rightarrow^3F_{2,3}$ and $^3H_4\rightarrow^1G_4$ are symbolized as $W_{25}$ and $W_{46}$, respectively, and $\sigma_{25}$ and $\sigma_{46}$ are the corresponding absorption cross sections. $k_{4267}$, $k_{6268}$, $k_{2464}$, and $k_{1463}$ are the coefficients of the CR and ETU process $^1G_4\rightarrow^1D_2\cdots^3H_4\rightarrow^3F_4$, $^3F_4\leftarrow^1G_4\rightarrow^3P_0$, $^3F_4\rightarrow^3H_4\leftarrow^1G_4$, and $^3H_6\rightarrow^3H_4\cdots^1G_4\rightarrow^3H_5$, respectively.
\subsubsection{$^3H_6\rightarrow^3F_4$}
\begin{align}
\frac{\partial N_2}{\partial t}&=W_{12}N_1-W_{21}N_2-\frac{N_2}{\tau_2},\\
N_1&=N_{0}-N_2,\\
\pm \frac{\partial P^{\pm}_p}{\partial z}&+\frac{\partial P^{\pm}_p}{v_g\partial t}=\Gamma_p[\sigma_{21}N_2-\sigma_{12}N_1]P^{\pm}_p-\alpha_pP^{\pm}_p,\\
\pm \frac{\partial P^{\pm}_s}{\partial z}&+\frac{\partial P^{\pm}_s}{v_g\partial t}=\Gamma_s\sigma_{21}N_2P^{\pm}_s-\alpha_sP^{\pm}_s+\Gamma_s\frac{2hc^2}{\lambda_s^3}\sigma_{21}\nonumber\\
&\times N_2\Delta\lambda_{ASE}.
\end{align}
The stimulated absorption and emission coefficients mentioned above can be described as:
\begin{equation}
W_{ij}=\frac{\Gamma_p\sigma_{ij}[P_p^{+}+P_p^{-}]\lambda_p}{hcA_{core}}\ \ ij=12,13,14,25,46,
\end{equation}
\begin{equation}
W_{ij}=\frac{\Gamma_s\sigma_{ij}[P_s^{+}+P_s^{-}]\lambda_s}{hcA_{core}}\ \ ij=34,21,
\end{equation}
where $\lambda_p$ is the central wavelength of the pump. Note that we ignore the stimulated emission at the pump wavelength and the stimulated absorption at the laser wavelength, except the cases under the $^3H_6\rightarrow^3F_4$ transition, because the stimulated emissions at these pump wavelengths are significant. So for this transition,
\begin{equation}
W_{21}=\frac{\Gamma_s\sigma_{21}[P_s^{+}+P_s^{-}]\lambda_s}{hcA_{core}}+\frac{\Gamma_p\sigma_{21}[P_p^{+}+P_p^{-}]\lambda_p}{hcA_{core}}.
\end{equation}
The boundary conditions of these equations are the same as that in \cite{yang14}, except that a CW pump is used here.
\subsection{Thermal model}
In the TDFLs, despite the pump sources, most of the heat is generated from the absorption of the pump power. The heat at splicing points or fiber facets due to the Fresnel reflection and scattering are comparably small and can be dissipated effectively by cooling apparatuses, such as V-grooves. So our thermal model is aimed to describe the temperature distribution $T$ of the gain fiber, which  can be described by the thermal conduction equation \cite{brown01},
\begin{equation}
\frac{1}{r}\frac{\partial}{\partial r}\Big[r\frac{\partial T}{\partial r}\Big]=-\frac{Q}{k}.
\end{equation}
where $Q$ is the heat power density. $k$ is the thermal conductivity. $r$ is the radial axis. This equation can be solved analytically by combining with the boundary conditions that describe double-clad fibers and cooling methods \cite{fan11}. So the temperature distribution can be expressed as
\subsubsection{Core}
\begin{align}
T_0(r)&=T_c+\frac{Qr_0^2}{2}\Big[\frac{1}{Hr_2}+\frac{1}{k_1}\ln(\frac{r_1}{r_0})+\frac{1}{k_2}\ln(\frac{r_2}{r_1})\nonumber\\
&+\frac{(1-r^2/r_0^2)}{2k_0}\Big],
\end{align}
where $T_c$ is the temperature of heat sink. $r_0$, $r_1$, $r_2$ are the radii of fiber core, inner cladding, and coating, respectively, and $k_0$, $k_1$, and $k_2$ are the corresponding thermal conductivity of materials. $H$ is the convective cooling coefficient.
\subsubsection{Inner cladding}
\begin{align}
T_1(r)=T_c+\frac{Qr_0^2}{2}\Big[\frac{1}{Hr_2}+\frac{1}{k_1}\ln(\frac{r_1}{r})+\frac{1}{k_2}\ln(\frac{r_2}{r_1})\Big],
\end{align}
\subsubsection{Coating}
\begin{align}
T_2(r)=T_c+\frac{Qr_0^2}{2}\Big[\frac{1}{Hr_2}+\frac{1}{k_2}\ln(\frac{r_2}{r})\Big].
\end{align}
We assume the heat generated is uniformly distributed in fiber core, So the heat power density at $z$ can be derived:
\begin{equation}
Q(z)=\frac{P_{absorb}(z)\eta_{heat}}{A_{core}\Delta z},
\end{equation}
where $P_{absorb}(z)$ is the absorbed pump power at $z$. $\Delta z$ is the numerical segment along the fiber axis. The conversion coefficient for the absorbed pump power to the heat $\eta_{heat}=1-\eta_{laser}$ we employed is the same as that in \cite{dawson08} and different from most of the previous works \cite{brown01,wang04,fan11}, in where only quantum defect were considered and thus the heat was underestimated.
\subsection{Parameters and numerical methods}
\begin{table}[tbh]
\caption{Parameters Employed in the Numerical Simulation}
\begin{center}
\begin{tabular}{ccc}
\hline
\hline
Parameter & Value & Ref.\\
\hline
$\sigma_{21}$ & $3.3\times 10^{-25}$ m$^2$ & \cite{jackson99} \\
$\sigma_{25}$ & $2.4\times 10^{-25}$ m$^2$ & \cite{jackson99}\\
$\sigma_{46}$ & $3.0\times 10^{-26}$ m$^2$& \cite{jackson98} \\
$\sigma_{34}$ & $5.5 \times 10^{-25}$ m$^2$& \cite{tang11}\\
$k_{1242}$ & $1.8\times 10^{-22}$ m$^3$s$^{-1}$ &\cite{tang11} \\
$k_{2124}$ & $1.5\times 10^{-23}$ m$^3$s$^{-1}$ & \cite{tang11}\\
$k_{2123}$ & $1.5\times 10^{-24}$ m$^3$s$^{-1}$ & \cite{tang11} \\
$k_{1232}$ & $3.0\times 10^{-24}$m$^3$s$^{-1}$ &  \cite{jackson99}\\
$k_{4267}$ & $6.0\times 10^{-23}$ m$^3$s$^{-1}$ & \cite{jackson99} \\
$k_{6268}$ & $1.2\times 10^{-23}$ m$^3$s$^{-1}$ & \cite{jackson99}\\
$k_{2464}$ & $1.6\times 10^{-23}$ m$^3$s$^{-1}$ &\cite{jackson99} \\
$k_{1463}$ & $1.0\times 10^{-23}$ m$^3$s$^{-1}$ & \cite{jackson99} \\
$\tau_2$ & 334.7 $\mu$s & \cite{jackson98} \\
$\tau_3$ & 0.007 $\mu$s & \cite{jackson98}\\
$\tau_4$ & 14.2 $\mu$s & \cite{jackson98} \\
$\tau_5$ & 0.0004 $\mu$s & \cite{jackson98}\\
$\tau_6$ & 783.8 $\mu$s & \cite{jackson98} \\
 $\tau_6$ & 783.8 $\mu$s & \cite{jackson98}\\
$\tau_7$ & 128.1 $\mu$s & \cite{jackson98} \\
$\tau_8$ & 284.9 $\mu$s & \cite{jackson98}\\
$T_c$ & 20 $^{\circ}$C & -\\
$k_0$, $k_1$ & 1.38 Wm$^{-1}$K$^{-1}$ & \cite{lapointe09} \\
$k_2$ & 0.24 Wm$^{-1}$K$^{-1}$ & \cite{lapointe09} \\
\hline \hline
\end{tabular}
\end{center}
\end{table}
Table \uppercase\expandafter{\romannumeral2} lists the parameters we employed to numerically solve the theoretical model. It is not a full list because some of them are varying in different analysis, such as the ion concentration $N_0$ and the gain fiber length. The model combining with the boundary conditions that describe the incident lights and the feedback of resonant cavity are then solved by a finite difference in time domain (FDTD) algorithm (see details in \cite{yang14} and references therein). The CW solutions are achieved when the variation of the output power less than $10^{-5}$.
\subsection{Comparison with experiments}
To check the validity of our model and numerical algorithm, we calculate the slope efficiency when the pump wavelengths in Section \uppercase\expandafter{\romannumeral2} are used, with the similar laser configurations and fiber parameters that were employed in the representative experiments \cite{ehrenreich10,hanna90,shen06,creeden14}.\\
\indent The calculated slope efficiency using different pump wavelengths are: 55.7\% (0.8 $\mu$m), 34.1\% (1 $\mu$m), 66.5\% (1.6 $\mu$m), and 92.3\% (1.9 $\mu$m). The discrepancy between the calculated efficiency and those from experiments is less than 8\%. Consider the amount of the parameters involved and the systemic errors in the experiments, it is an acceptable result for predicting the behaviors of the TDFLs at kilowatt level.
\section{Results}
In this Section, we employ the model to analysis the thermal behaviors of high-power TDFLs. Only cladding-pump scheme is used because it takes advantage of well-developed fiber pump combiners and a large pump area will significantly reduce the pump intensity and thus waken the effects of pump saturation \cite{sanchez98}. For consistency, we employ a TDF with a widely-used core/inner cladding diameter of 25/400 $\mu$m and a Tm$^{3+}$ doping concentration of $3.5\times10^{26}$ m$^{-3}$ for high efficiency of the ``two-for-one'' CR process. We use the perpendicularly-cleaved facet of fiber as the output coupler for high slope efficiency, corresponding to a reflectivity of 0.04. Due to the diversity of the pump absorption when different pump transitions are employed, the lengths of the TDFs are varying. In the simulation, we adopt the gain fiber lengths when the maximum slope efficiencies are achieved.
 \subsection{Power scaling}
\begin{figure}[htb]
\begin{center}
\includegraphics[scale=0.3]{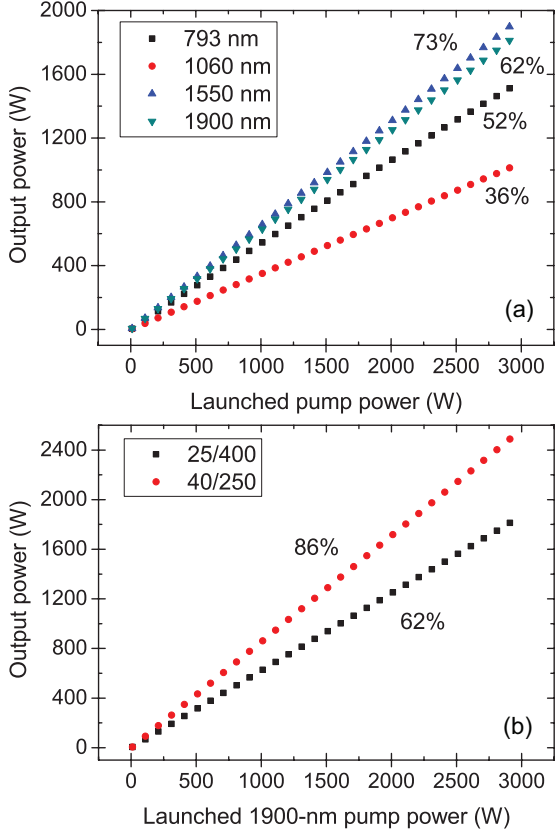}
\caption{(a) Power scalings using the widely-used 25/400 TDF. (b) Comparison of laser performances using 25/400 and 40/250 TDFs pumped at 1900 nm.}
\end{center}
\end{figure}
We first calculate the lasing performances under different pump transitions. The results are shown in Fig.~2(a). The slope efficiencies pumped at 793 nm, 1060 nm, and 1550 nm are similar to that retrieved in the experiments, are 52\%, 36\%, 73\%, respectively. But the slope efficiency pumped at 1900 nm is only 62\%, much lower than the experimental one. The decrease of the slope efficiency is resulted from the small power filling factor of this kind of fiber (0.004), which largely reduces the absorption of the pump power and thus a long gain fiber must be used for optimal efficiency and thus accumulate the instinct losses both at the pump (1.9 $\mu$m) and the laser (2 $\mu$m) wavelengths..\\
\indent We thus employ a TDF with a core/inner cladding diameter of 40/250 $\mu$m, which leads to the increase of the power filling factor to $0.02$ for cladding-pump. Fiber with such geometry has been fabricated to develop high-power cladding-pumped holmium-doped fiber lasers \cite{hemming13}. Fig.~2(b) shows the laser performances under different fiber geometry. A slope efficiency of 86\% is achieved when the 40/250 TDF is used. By increasing the power filling factor, the slope efficiency of the laser can be further improved. We will discuss it in Section \uppercase\expandafter{\romannumeral5}.
\subsection{Influences of pump}
The temperature raises of the gain fiber caused by pump absorption are investigated. A convective cooling coefficient $H$ of 1000 Wm$^{-1}$K$^{-1}$ is adopted in this subsection. Note that we only pay attention to the position having the maximum temperature when a specific level of pump power is injected, because the TDFLs will be immediately destroyed due to fiber failure at this position. In most experimental conditions (also assumed in this work), the convection cooling along the gain fiber is homogeneous, so the maximum temperature should happen at $z=0$. The fiber failure, on the other hand, includes the damage of fiber core and the burning of fiber coating. The former related to a critical temperature of 700 $^{\circ}$C \cite{hand88}, where the fiber fuse effect happens and the later related to the damage temperature of the polymer coating of 200 $^{\circ}$C \cite{jeong08}. So we concentrate on two kinds of temperature. 1) At $(z=0,r=0)$, where has the highest core temperature. 2) At $(z=0,r=r_1)$, where has the highest coating temperature. We define them as the core temperature and the coating temperature, respectively. \\ 
\begin{figure}[htb]
\begin{center}
\includegraphics[scale=0.3]{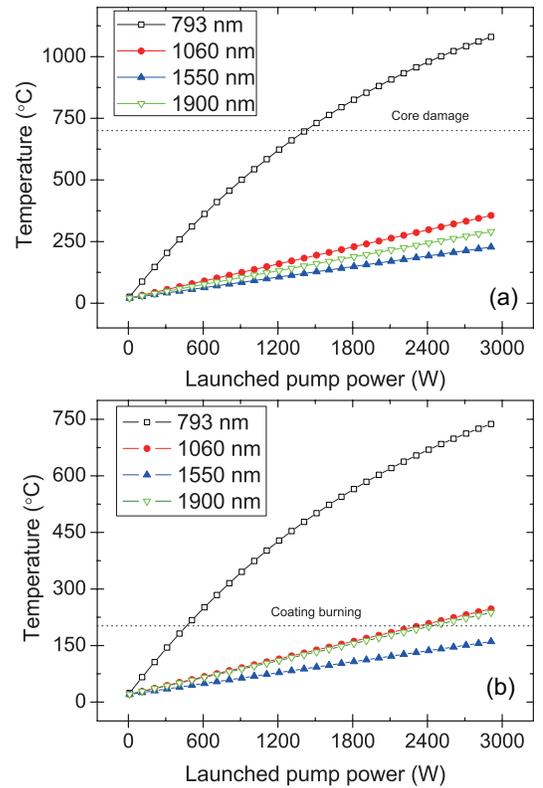}
\caption{Core (a) and coating (b) temperature as a function of launched pump power when different pump wavelengths are used.}
\end{center}
\end{figure}
\indent Figure~3 demonstrates the core (a) and coating (b) temperature as a function of launched pump power when different pump wavelengths are used. The temperature increases as more pump power is launched. Except the one pumped at 793 nm, the temperature increases of the core (coating) are linear with a rate of 0.12 (0.08), 0.07 (0.05), 0.09 (0.07) $^{\circ}$C/W for those pumped at 1060 nm, 1550 nm, 1900 nm, respectively. Because the 40/250 TDF is employ when pumped at 1900 nm, it large absorption leads to a quicker temperature rise than that pumped at 1550 nm. Even so, the thermal performance of the TDFLs pumped at these wavelength are similar. On the other hand, roll-over of the temperature rises is observed when pumped at 793 nm. We find it is the saturation of pump absorption caused by the combination of the high Tm$^{3+}$ concentration we employed and the large absorption cross section of the Tm$^{3+}$ cations. Despite this, we acquire an average rate of the temperature rises of 0.36 $^{\circ}$C/W for the core and 0.24 $^{\circ}$C/W for the coating, which is significantly larger than those pump at the long wavelengths. From the perspective of the fiber failure, the coating burning and the core damage happen when the launched 793-nm pump power are 460 W and 1420 W, respectively, which correspond to 2-$\mu$m laser powers of 241 W and 756 W. While no concern of the core damage should be put on the kilowatt TDFLs pumped at 1060 nm, 1550 nm, and 1900 nm and the coating burning only happens when the output 2-$\mu$m powers close to or exceed 1 kW. \\
\begin{figure}[htb]
\begin{center}
\includegraphics[scale=0.3]{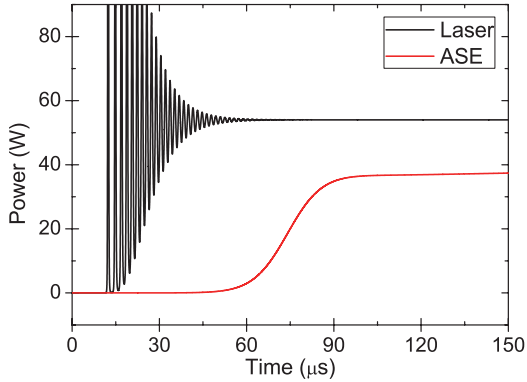}
\caption{Temporal evolution of the laser and the ASE in the numerical simulation.}
\end{center}
\end{figure}
\indent Comparing to directly output kilowatt lasers from a laser oscillator, MOPA configurations are usually adopted to construct high-power TDFLs. The extra problem related to fiber amplifiers is the portion of heat resulting from insufficient extraction of pump energy, the amount of which relates to the level of injected seed. The upper limit of this potion of heat happens when the lasers only output ASE, namely the superfluorescent sources \cite{shen08}. We realize this operation in the simulation by setting the feedback of the laser cavity to be 0. Figure~4 shows the temporal evolution of the ASE. The laser operation is also plotted for comparison. The differences between them are the relaxation oscillation at the beginning of laser formation and the output levels when reach a steady-state. Note that the ASE in the figure is the sum of the outputs from both ends of the gain fiber. For the TDFLs pumped at 793 nm, 1060 nm, 1550 nm, and 1900 nm, the slope efficiency of the ASE operation are 38\%, 23\%, 53\%, 59\%, respectively.\\
\begin{figure}[htb]
\begin{center}
\includegraphics[scale=0.18]{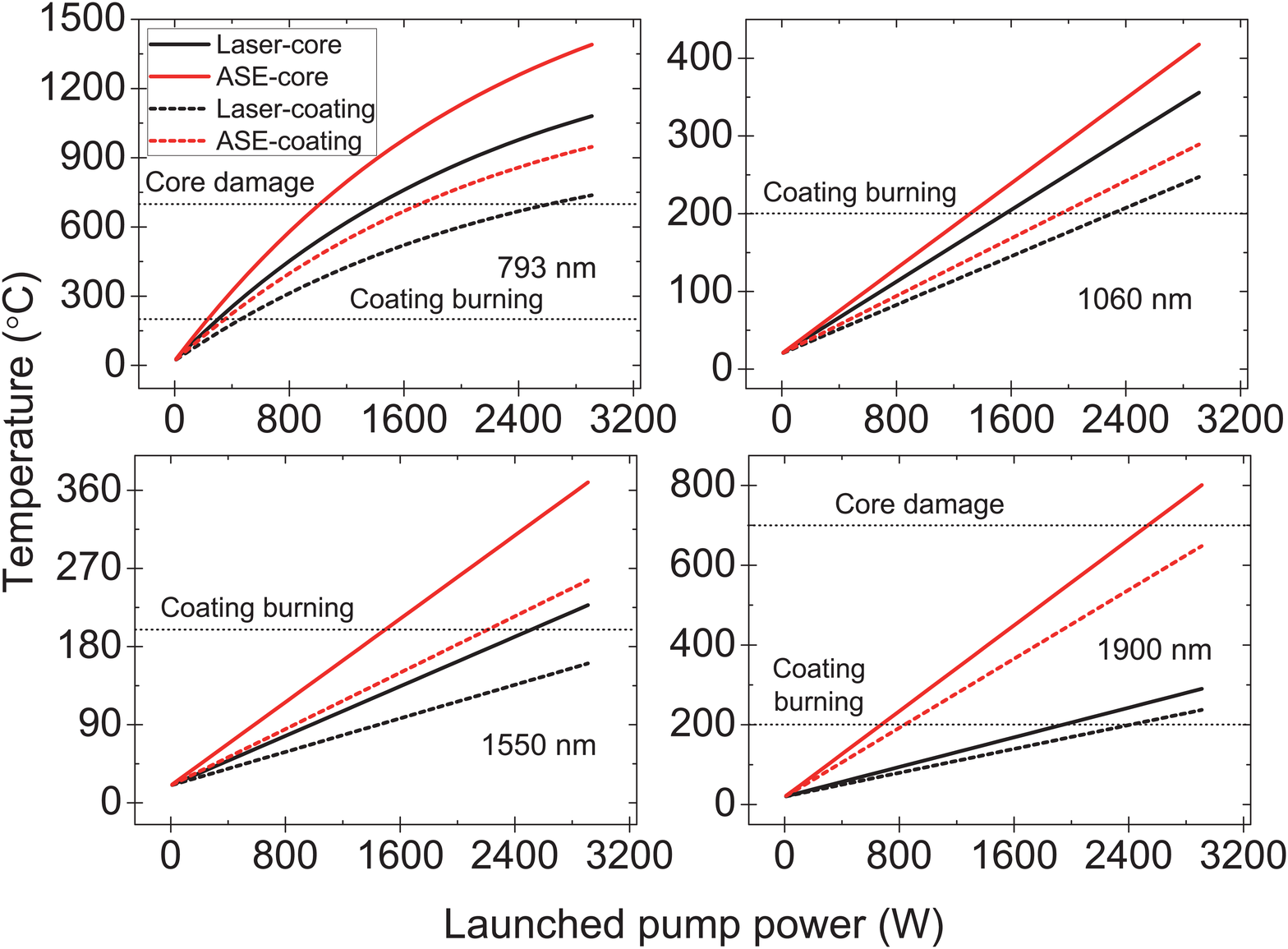}
\caption{Core and coating temperatures as a function of launched pump power for both the laser and the ASE operations.}
\end{center}
\end{figure}
\indent Figure~5 demonstrates the core and coating temperatures as a function of launched pump power for both the laser and the ASE operations. The real operation states of thulium-doped fiber amplifiers (TDFAs) fall in between these two operations. For the TDFLs pumped at 793 nm, the rise rate of the temperature of the core (coating) increases from 0.36 (0.24) $^{\circ}$C/W for the laser operation to 0.46 (0.31) $^{\circ}$C/W for the ASE operation, which results in the launched pump powers causing the core damage and the coating burning decrease from 1420 W and 460 W to 1020 W and 360 W. The situations are similar for the TDFLs pumped at 1060 nm and 1550 nm. The increases of the rate are 0.02 and 0.04 $^{\circ}$C/W, respectively, for both the core and the coating. It implies no significant variation of the thermal features happening from the TDFLs to the TDFAs for these pump wavelengths. As to the TDFLs pumped at 1900 nm, rapid increases of the temperature rise rate are observed. Their rise rates of the core (coating) increases from 0.09 (0.07) $^{\circ}$C/W for the laser operation to 0.27 (0.22) $^{\circ}$C/W for the ASE operation, which leads to the launched pump power inducing the coating burning reduces from 2400 W to 840 W and that inducing the core damage drops to 2540 W. The large transition of the thermal characteristic from the laser operation to the ASE operation is resulted from a larger change of the slope efficiency (86\% $\rightarrow$ 59\%) comparing those pumped at other wavelengths (52\% $\rightarrow$ 38\% at 793 nm; 36\% $\rightarrow$ 23\% at 1060 nm; 73\% $\rightarrow$ 53\% at 1550 nm.).
\subsection{Convective cooling}
\begin{figure}[htb]
\begin{center}
\includegraphics[scale=0.18]{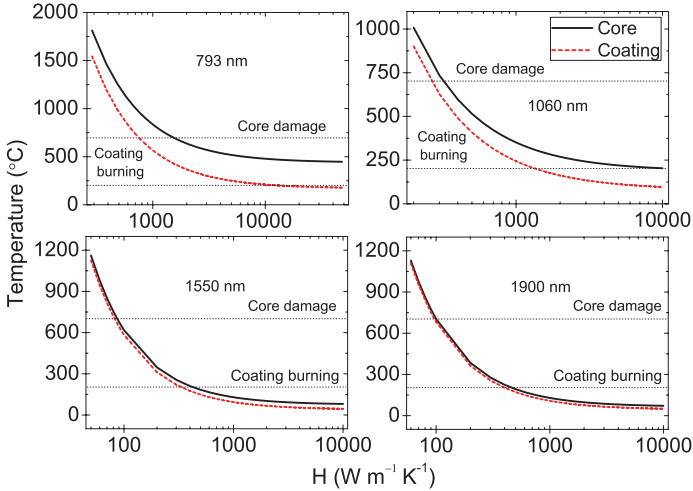}
\caption{Temperature as a function of the convective cooling coefficient $H$ when the TDFLs pumped at different wavelengths output 1-kW laser power.}
\end{center}
\end{figure}
Active cooling is important for the heat removal of high-power TDFLs, we investigate the influence of different convective cooling coefficients $H$ on the temperature of the TDFLs outputting 1-kW powers as demonstrated in Fig.~6. As the $H$ increase, the temperatures of the TDFLs pumped by all of the wavelengths drop. However, to keep the safety of the core and the coating, different levels of the $H$ are needed for different pump wavelengths. For the TDFLs pumped at 1550 nm and 1900 nm, A $H$ of less than 400 Wm$^{-1}$K$^{-1}$ can avoid the fiber failure. This value increases to 1320 Wm$^{-1}$K$^{-1}$ for those pumped at 1060 nm. For the well-used 793-nm pump, A $H$ of 11810 Wm$^{-1}$K$^{-1}$ must be employed to keep the safety of the gain fiber, which is difficult to realize based on the state-of-the-art fiber cooling techniques \cite{fan11,lapointe09}.
\section{Discussion}
In this Section, we propose the tandem-pumped TDFLs and discuss their feasibility.
\subsection{Laser configuration}
Based on the analysis above, using the pump lights at 1060 nm, 1550 nm, and 1900 nm would significantly decrease the temperature of the gain fiber, comparing to the TDFLs pumped at 793 nm. Without the consideration of cost and efficiency, they are all promising candidates. As shown in Fig.~7, taking advantage of the pump combiner, several YDFLs/EYFLs/TDFLs with tens-of-watts outputs can be efficiently coupled to pump the TDFLs and dual-directional pump would further reduce the temperature by a factor of two. Besides, the high-brightness of the fiber lasers would eliminate the power scaling limit induced by pump brightness \cite{dawson08}.\\
\begin{figure}[htb]
\centering
\includegraphics[scale=0.35]{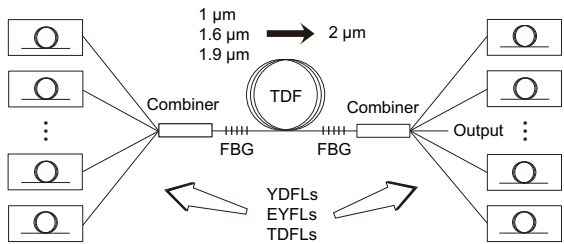}
\caption{Configuration of tandem-pumped high-power TDFLs}
\end{figure}
\indent However, employing the tandem-pumping strategy (LDs$\rightarrow$YDFLs/EYFLs/TDFLs$\rightarrow$TDFLs) would be inevitably at the cost of wall-plug efficiency, which may be ignored in laboratory but must be considered when goes to application and commercialization. The conversion efficiency at different stages of the tandem-pumped TDFLs is shown in Table \uppercase\expandafter{\romannumeral3}. 
\begin{table}[htb]
\caption{Efficiency of different tandem-pumping strategies}
\begin{center}
\begin{tabular}{c|ccc}
\hline
\hline
\multirow{2}{*}{Efficiency (\%)} &\multicolumn{3}{c}{Pump wavelength} \\
\cline{2-4}
&1060 nm & 1550 nm & 1900 nm\\
\hline
Electric power$\rightarrow$LDs & 50 & 50 & 50\\
LDs$\rightarrow$YDFLs/EYFLs/TDFLs& 83 & 40 & 53\\
YDFLs/EYFLs/TDFLs$\rightarrow$TDFLs&37& 72 & 90\\
\hline
\textbf{Wall-plug}&\textbf{15}&\textbf{14}&\textbf{24}\\
\hline
\hline
\end{tabular}
\end{center}
\end{table}
We assume a conversion efficiency from electric power to light of 50\%  for both 79x-nm and 9xx-nm LDs. The slope efficiency of LD-pumped YDFLs, EYFLs, and TDFLs are 83\%, 40\%, and 53\%, respectively, taken from representative experimental works \cite{jeong04,jeong07ey,ehrenreich10}. The slope efficiency of fiber-laser-pumped TDFLs has been stated above. Combining all of them, the wall-plug efficiency of the pump fiber lasers at 1060 nm, 1550 nm, and 1900 nm are 15\%, 14\%, and 24\%, respectively. The high wall-plug efficiency of the LDs$\rightarrow$TDFLs$\rightarrow$TDFLs make it advantageous over other pump strategies and this efficiency is comparable to available commercial TDFLs and solid-state counterparts.
\subsection{Fibers for long-wavelength cladding-pump}
As demonstrated in Section \uppercase\expandafter{\romannumeral4}, changing the geometry of the gain fiber from 25/400 to 40/250 can significantly improve the slope efficiency of the TDFLs pumped at 1900 nm. We further investigate this phenomenon by varying the diameter of the inner cladding to see how the slope efficiency and the optimized gain fiber length change. The core diameter is fixed at 40 $\mu$m.\\
\begin{figure}[htb]
\begin{center}
\includegraphics[scale=0.32]{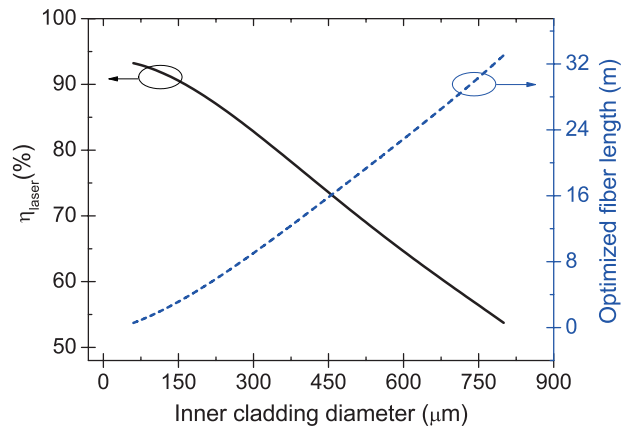}
\caption{Slope efficiency and optimized gain fiber length as a function of inner cladding diameter using the 1900-nm pump.}
\end{center}
\end{figure}
\indent In Fig.~8, as the diameter of the inner cladding increases, the optimized length of the gain fiber increases while the slope efficiency decreases. With a diameter of 800 $\mu$m, corresponding to a power filling factor of 0.0025, the slope efficiency is dropped to 53.7\% and the optimized gain fiber length is 33.5 m. Such a long fiber will not only increase the propagation loss of the laser and also accumulate nonlinear effects, such as stimulated Raman scattering, which are unwanted for power scaling \cite{dawson08}. On the other hand, with an inner cladding diameter of 60 $\mu$m, corresponding to a power filling factor of 0.44, the slope efficiency is 93.2\% close to the quantum efficiency (95\%) and the optimal length of the gain fiber is 0.6 m. There parameters are favorable for the operation of multi-kilowatt TDFLs. Besides, using the high-brightness fiber lasers as the pump avoids the demand of a high NA inner cladding, which brings convenient to the design of small inner cladding fibers.\\
\begin{figure}[htb]
\begin{center}
\includegraphics[scale=0.4]{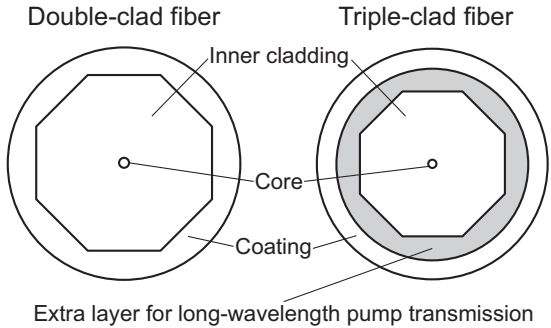}
\caption{Cross sections of double- and triple-clad fibers.}
\end{center}
\end{figure}
\indent To realize the tandem-pumped TDFLs we proposed, another obstacle is that the polymer coating of double-clad fiber will absorb the lights at wavelengths longer than 1 $\mu$m \cite{creeden14}, which will result in the coating burning when certain amount of long-wavelength pump power is launched. A straight-forward solution is to replace the polymer with a material which is transparent to the long-wavelength pump. One of this kind fiber has been demonstrated in \cite{hemming13} very recently. The researchers used fluoride glass to replace polymer. To keep mechanical strength and avoid surface damage of fiber, the polymer coating was still employed outside the fluoride layer. Comparing to traditional double-clad fibers, an extra layer for long-wavelength pump transmission make the new fiber triple-cladding as demonstrated in Fig.~9. Beyond this method, other types of fibers such as photonic crystal fibers may also be promising for guiding long-wavelength pump lights.\\
\begin{figure}[htb]
\begin{center}
\includegraphics[scale=0.32]{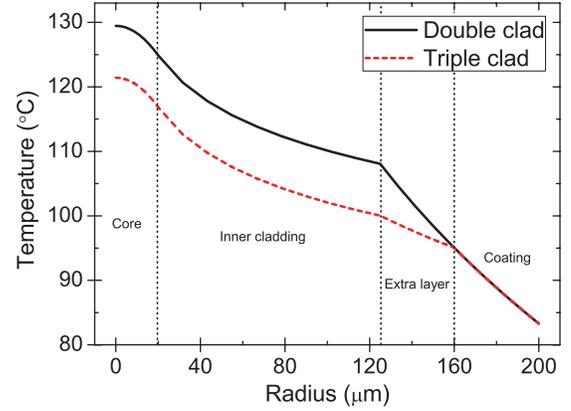}
\caption{Radial distributions of temperature for double- and triple-clad fibers.}
\end{center}
\end{figure}
\indent We further investigate the thermal performance of the triple-clad fiber. We employ the fluoride glass to constitute the extra layer for example, which has a thermal conductivity of 0.628 Wm$^{-1}$K$^{-1}$ \cite{zhu10}. The core/inner cladding/coating diameter of the double-clad fiber we used is 40/250/400 $\mu$m. For the triple-clad fiber, we employ the same geometry to keep the lasing performance unchanged and substitute a part of the coating with the fluoride glass, which has a thickness of 70 $\mu$m. Figure~10 shows the radial distributions of temperatures for the double- and triple-clad fibers when a TDFL pumped at 1900 nm outputs 1-kW 2-$\mu$m laser at a $H$ of 1000 Wm$^{-1}$K$^{-1}$. Due to the fluoride glass has a larger thermal conductivity than the polymer, which leads to a more efficient heat dissipation, the temperature of the triple-clad fiber in the core and inner cladding is 8 $^{\circ}$C lower than that of the double-clad fiber. So we can conclude that adding a extra layer which has a larger thermal conductivity than the polymer will be favorable for the heat dissipation of fiber.
\section{Conclusion}
Based on an experimental-validated model, we have numerically analyzed thermal characteristics of multi-kilowatt TDFLs using different pump transitions. our results show the temperature of the gain fiber will be significantly decreased by employing the pump transition $^3H_6\rightarrow^3F_4$ and $^3H_6\rightarrow^3H_5$, comparing to the TDFLs pumped by high-power LDs at 79x nm ($^3H_6\rightarrow^3H_4$). We have thus developed the tandem-pumped TDFLs and discussed their feasibility and design. We have also analyzed the temperature of the gain fiber under different convection cooling conditions and the difference between the laser and superfluorescent operations. The thermal analysis in this work can be referenced in the design of high-power TDFLs and TDFAs. The proposed tandem-pumping strategy is promising to achieve multi-kilowatt 2-$\mu$m lasers with low thermal load.
\ifCLASSOPTIONcaptionsoff
  \newpage
\fi

\end{document}